\documentclass[preprint,showpacs,preprintnumbers,amsmath,amssymb]{revtex4-1}

\usepackage{graphicx}
\usepackage{dcolumn}
\usepackage{bm}

\begin{document}


\title{Transient behaviour of a polymer dragged through a viscoelastic medium}
\author{Hans Vandebroek$^1$}
\author{Carlo Vanderzande$^{1,2}$}
\affiliation{%
$^1$Faculty of Sciences, Hasselt University, 3590 Diepenbeek, Belgium. \\
$^2$Instituut Theoretische Fysica, KULeuven, 3001 Heverlee, Belgium.
} 


\date{\today}

\begin{abstract}
We study the dynamics of a polymer that is pulled by a constant force through a viscoelastic medium. This is a model for a polymer being pulled through a cell by an external force, or for an active biopolymer moving due to a self generated force. Using a Rouse model with a memory dependent drag force, we find that the center of mass of the polymer follows a subballistic motion.  We determine the time evolution of the length and the shape of the polymer. Through an analysis of the velocity of the monomers, we investigate how the tension propagates through the polymer. We discuss how polymers can be used as a probe of the properties of a viscoelastic medium.

 \end{abstract}

\pacs{82.35.Lr, 36.20.Ey, 05.10.Gg}
\maketitle
\section{Introduction}
While the static and dynamic properties of polymers in equilibrium are rather well understood \cite{Doi86}, much less is known about the behaviour of polymers in nonequilibrium situations. One of the simplest cases that comes to mind is that of a polymer that is pulled by a constant force through a medium or the related problem of a polymer that is held fixed in a uniform flow. Scaling arguments showed that for polymers with a finite extensibility,  a trumpet shape is assumed when the applied forces or fluid velocities are small \cite{Brochard93}, whereas when the forces are sufficiently strong a stem-flower shape appears \cite{Brochard95}. We are not aware of experiments in which a polymer is dragged through a medium, though experiments in which a fluorescent DNA is held fixed in a uniform flow using an optical tweezer have been performed \cite{Perkins95,Larson97}. More recently, also the transient response of a polymer to an applied force was investigated theoretically \cite{Rowghanian12, Sakaue12}. On the basis of exact calculations in the Rouse model and scaling arguments, it was found that after a force is suddenly applied to one end of a polymer, a tension front propagates towards the other end of the polymer such that the $n$-th monomer is set into motion in a time that grows as $n^2$. It has been argued that tension propagation also plays an essential role in polymer translocation \cite{Sakaue10, Rowghanian11, Panja08}. 

In cellular biophysics, several situations are known where a polymer is being dragged through the cytoplasm of the cell:  DNA that is pulled by the mitotic spindle during cell division \cite{Boal12}, or forces applied by the actin-myosin complex on immersed microtubuli \cite{Brangwynne08a} . 
Or we can think of active polymers like molecular motors that use the energy liberated from ATP hydrolysis to generate their own force to transport various cargos across the cell \cite{Howard01}. These cases differ from those studied in \cite{Brochard93,Brochard95,Sakaue12,Rowghanian12} in that the cell is a complex, crowded medium that can to a good approximation be described as a viscoelastic fluid. 

This then naturally leads to the question of how a polymer reacts to an applied force (self generated or due to the environment) in a viscoelastic medium. Besides being an interesting question for polymer physics, the response can also be used to obtain properties of the medium itself. Rheological properties of the cytoplasm have been obtained mostly by studying the response, or the related fluctuations of small objects (polymers, vesicles, nanoparticles, ...)\cite{Lau03,Robert10}. These objects were mostly considered as point particles, by, for example, looking at the diffusive behaviour of their center of mass \cite{Hofling13}. However, polymers are extended objects and by also looking at their geometric properties, additional insight on the flow properties of the cytoplasm can be obtained. 

In the present paper we study an extremely simplified, analytically tractable version of that problem. The results obtained here may give some indications of what to expect in more realistic models where one will have to use approximate calculational techniques or simulations.

The model we use for the polymer is that of a Rouse chain. In a viscoelastic medium \cite{Doi86}, drag forces have a memory of the past, so the usual Stokesian friction force $-\gamma v(t)$ on a particle is replaced by an integral with a kernel $K(t)$: $-\gamma \int K(t-t') v(t') dt'$. The kernel has a power law decay $K(t) \sim t^{-\alpha}$. The value of the exponent $\alpha$ can be determined both from active and passive rheological measurements and estimates for cellular environments range from $\alpha \approx 0.1$ to $0.4$ \cite{Fabry01,Robert10,Weber12}. This defines the model we use. Our calculations extend earlier ones that investigated the diffusion of a Rouse chain in a viscoelastic medium in absence of a force \cite{Weber10}. The results obtained were used to analyse the subdiffusive motion of chromosomal loci in bacteria and yeast \cite{Weber12}.

We will focus on the time evolution of geometric properties of the polymer such as its density, its average length and its fluctuations. Besides this, we will also investigate the tension propagation through the chain.

This paper is organised as follows. In section II  we introduce the model and its description in terms of eigenmodes. In section III, we look at the average motion of the center of mass while in section IV we look at average properties of individual monomers. This allows us to determine the shape of the polymer in section V and the propagation of tension through the polymer in section VI. In section VII we look at fluctuating quantities, like the diffusion of the center of mass and fluctuations in polymer length. Finally in section VIII we present some reflections on the experimental realisation of our theory and give some further conclusions. 

\section{The Rouse model in a viscoelastic medium} 
In the standard Rouse model, a polymer with $N$ monomers is modelled as a system of $N$ beads connected by springs. Immersed in a solvent, each monomer feels moreover a friction and a random force. The position of the $n$-th monomer, $\vec{R}_n(t)$, then obeys a Langevin equation
\begin{eqnarray}
\gamma \frac{d\vec{R}_n (t)}{dt} = - k\left(2 \vec{R}_n(t) - \vec{R}_{n+1}(t) - \vec{R}_{n-1}(t)\right) + \vec{F}_n(t)
\label{1}
\end{eqnarray}
where $\gamma$ is the friction coefficient, and $k$ the spring constant. The force $\vec{F}_n(t)$ is the sum of a deterministic force $\vec{f_n}(t)$ and a random force $\vec{\xi}_n(t)$: $\vec{F}_n(t) = \vec{f}_n(t) + \vec\xi_n(t)$. The random force $\vec{\xi}_n(t)$ is a Gaussian random process whose covariance is given by the fluctuation-dissipation theorem: $\langle \vec{\xi}_n(t)\cdot \vec{\xi}_m(t') \rangle =6 \gamma k_B T \delta(t-t') \delta_{nm}$.

In the present case, the solvent is considered to be a viscoelastic medium in which friction becomes memory dependent: $\gamma \int_0^t K(t-\tau) (d\vec{R}_n(\tau)/dt) d\tau$. A good description of a viscoelastic medium is obtained by assuming a power law for the kernel. We choose the parametrisation
\begin{eqnarray}
K(t) = \frac{(2-\alpha)(1-\alpha)}{|t|^\alpha}
\label{2}
\end{eqnarray}
In this way, viscous behaviour is recovered for $\alpha=1$. For $\alpha=0$ an elastic response is recovered. Hence for $0<\alpha<1$, we have the viscoelastic case, intermediate between the elastic and viscous response. The equation of motion for the $n$-th monomer now becomes the generalised Langevin equation
\begin{eqnarray}
\gamma \int_0^t K(t-\tau)\frac{d\vec{R}_n (\tau)}{dt}d\tau=- k\left(2 \vec{R}_n(t) - \vec{R}_{n+1}(t) - \vec{R}_{n-1}(t)\right) +  \vec{F}_n(t)
\label{3}
\end{eqnarray}
The random force remains a Gaussian process with average $\langle \vec{\xi}_n(t)\rangle=0$ and a correlation function that is connected to the kernel $K(t)$ by the fluctuation-dissipation theorem
\begin{eqnarray}
\langle \vec{\xi}_n(t)\cdot \vec{\xi}_m(t') \rangle = 3\gamma k_B T K(|t-t'|) \delta_{nm}
\label{4}
\end{eqnarray}
Mathematically, for the kernel (\ref{2}), $\xi(t)$ becomes so called fractional Gaussian noise \cite{Qian03}.
In the present paper, $\vec{f}_n(t)$ is a constant force that is turned on at $t=0$ and that acts only on the first monomer (we label the monomers from $n=0$ to $N-1$)
\begin{eqnarray}
\vec{f}_n(t) = f \delta_{n0} H(t) \hat{e}_x
\label{5}
\end{eqnarray}
where $H(t)$ is the Heaviside function. We assume that at $t=0$ the polymer is in thermal equilibrium. Finally, we introduce ghost monomers with $n=-1$ and $n=N$ whose positions obey $\vec{R}_{-1}(t)=\vec{R}_0(t)$ and $\vec{R}_N(t)=\vec{R}_{N-1}(t)$ so that all physical monomers obey (\ref{3}). The solution of (\ref{3}) with these boundary conditions and for $f=0$ was discussed in \cite{Weber10} in the context of the motion of chromosome loci.

In order to solve (\ref{3}), it is common to introduce the Rouse modes
\begin{eqnarray}
\vec{X}_p(t) = \frac{1}{N} \sum_{n=0}^{N-1} \vec{R}_n(t) \cos \left( \frac{\pi p}{N} (n+{}^1\!/_2)\right)
\label{6}
\end{eqnarray}
for $p=0,\ldots,N-1$. 
A straightforward calculation shows that $\vec{X}_p(t)$ obeys the generalised Langevin equation
\begin{eqnarray}
\int_0^t K(t-\tau) \frac{d\vec{X}_p (\tau)}{dt} d\tau = - \frac{\vec{X}_p(t)}{\tau_p} + \frac{\vec{F}_p(t)}{\gamma}
\label{7}
\end{eqnarray}
Here $\tau_p$ is the relaxation time of the $p$-th mode
\begin{eqnarray}
\tau_p = \frac{\gamma}{4k \sin^2(\pi p/2 N)}
\label{8}
\end{eqnarray}
which for $N$ large can be approximated as $\tau_p \approx \gamma N^2/p^2 \pi^2 k$. The force $\vec{F}_p(t)$ equals
\begin{eqnarray}
\vec{F}_p(t) = \frac{1}{N}  \sum_{n=0}^{N-1}  (\vec{f}_n(t) + \vec{\xi}_n(t)) \cos \left( \frac{\pi p}{N} (n+{}^1\!/_2)\right)
\label{9}
\end{eqnarray}
\section{The center of mass motion}
The normal mode with $p=0$ corresponds to the center of mass $\vec{R}_{cm}(t)=\sum_n \vec{R}_n(t)/N$. Its equation of motion
\begin{eqnarray}
\int_0^t K(t-\tau) \frac{d\vec{X}_0 (\tau)}{dt} d\tau = \frac{\vec{F}_0(t)}{\gamma}
\label{10}
\end{eqnarray}
can be solved by Laplace transform. The calculation is simple and gives
\begin{eqnarray}
\vec{X}_{0}(t) - \vec{X}_{0}(0) = \frac{1}{\gamma G_\alpha} \int_0^t \vec{F}_0(t-\tau) \tau^{\alpha-1} d \tau
\label{11}
\end{eqnarray}
where $G_\alpha=\Gamma(\alpha)\Gamma(3-\alpha)$. Averaging over different histories of the process, we obtain
\begin{eqnarray}
\langle \vec{R}_{cm}(t) \rangle =\langle \vec{X}_{0}(t) \rangle=  \frac{f}{\gamma N G_\alpha \alpha} t^{\alpha}\ \hat{e}_x
\label{12}
\end{eqnarray}
(taking $\langle\vec{R}_{cm}(0)\rangle=0$). This is our first conclusion: the centre of mass moves subballistically. Such motion has indeed been observed in experiments with molecular motors that move under their own, actively generated, force \cite{Caspi00}.

The average velocity of the center of mass decreases as a power law
\begin{eqnarray}
\langle \vec{V}_{cm}(t)\rangle = \frac{f}{\gamma N G_\alpha} t^{\alpha-1}\ \hat{e}_x
\label{13}
\end{eqnarray}
For the viscous case $\alpha=1$, we recover the result of \cite{Sakaue12}, $\langle \vec{V}_{cm}\rangle= (f/N\gamma)\hat{e}_x$. For $\alpha<1$, the polymer comes to rest for large $t$. 

\section{Position and velocity of the monomers}
We are now interested to calculate the (average) position and velocity for each of the monomers. This will allow us to determine the shape of the polymer as a function of time. Moreover, the results on the velocity of the monomer will give insight in how the force acting on the first monomer propagates through the polymer. 

As a first step, we have to solve (\ref{7}). This can again be done using Laplace transforms. The calculation is straightforward, details are given in Appendix A. The resulting time evolution of the modes can be written in terms of the generalised Mittag-Leffler functions $E_{\alpha,\beta}(z)$ \cite{Haubold11} which are defined as:
\begin{eqnarray}
E_{\alpha,\beta}(z) = \sum_{n=0}^\infty \frac{z^n}{\Gamma(\alpha n + \beta)}
\label{14}
\end{eqnarray}
The Mittag-Leffler function can be seen as an extension of the exponential function to which it reduces for $\alpha=\beta=1$. 
For $|z| \to \infty, \arg(z)=-\pi$, we have, for $\alpha \neq \beta$ the asymptotic behaviour
\begin{eqnarray}
E_{\alpha,\beta}(z) \simeq -\frac{z^{-1}}{\Gamma(\beta-\alpha)}
\label{15}
\end{eqnarray}
while for $\alpha=\beta$ 
\begin{eqnarray}
E_{\alpha,\beta}(z) \simeq -\frac{z^{-2}}{\Gamma(-\alpha)}
\label{15b}
\end{eqnarray}
Solving (\ref{7}), we find for $\vec{X}_p(t)$
\begin{eqnarray}
\vec{X}_p(t) = \vec{X}_p(0) E_{\alpha,1}\left(- \left(\frac{t}{\tau_{p,\alpha}}\right)^\alpha\right) + \frac{1}{\gamma \Gamma(3-\alpha)} \int_0^t \vec{F}_p(t-\tau) \tau^{\alpha-1} E_{\alpha,\alpha} \left(- \left(\frac{\tau}{\tau_{p,\alpha}}\right)^\alpha\right) d\tau \nonumber \\
\label{15c}
\end{eqnarray}
Here $\tau_{p,\alpha}$ is a set of time scales appearing in the problem
\begin{eqnarray}
\tau_{p,\alpha} = \left(\Gamma(3-\alpha) \tau_p \right)^{1/\alpha} 
\label{16}
\end{eqnarray}
For $N$ large, $\tau_{p,\alpha}  \sim \left(\frac{\gamma N^2}{p^2}\right)^{1/\alpha}$. 

We average (\ref{15c}) over initial positions $\langle\vec{X}_p(0)\rangle$ and histories using
\begin{eqnarray}
\langle \vec{F}_p(t) \rangle &=& \frac{1}{N}\sum_{n=0}^{N-1}  f\delta_{n,0} H(t) \hat{e}_x \cos\left(\frac{\pi p}{N}(n+{}^1\!/_2)\right)  \nonumber \\
&=& \frac{f}{N} \cos \left(\frac{\pi p}{2N}\right) H(t) \hat{e}_x
\label{17}
\end{eqnarray}
This gives 
\begin{eqnarray}
\langle \vec{X}_p(t)\rangle &=& \frac{f \cos \left(\frac{\pi p}{2N}\right)}{N \gamma \Gamma(3-\alpha)}  \int_0^t \tau^{\alpha-1} E_{\alpha,\alpha} \left(- \left(\frac{\tau}{\tau_{p,\alpha}}\right)^\alpha\right) d\tau \hat{e}_x\nonumber \\
&=& \frac{f \cos \left(\frac{\pi p}{2N}\right)}{N\gamma \Gamma(3-\alpha)} t^\alpha E_{\alpha,\alpha+1} \left(- \left(\frac{t}{\tau_{p,\alpha}}\right)^\alpha\right) \hat{e}_x
\label{18}
\end{eqnarray}

The position of the $n$-th monomer is then written in terms of the Rouse modes by inverting  (\ref{6}).  This gives 
\begin{eqnarray}
\vec{R}_n(t) = \vec{X}_0(t) + 2 \sum_{p=1}^{N-1} \vec{X}_p(t) \cos \left( \frac{\pi p}{N}(n+{}^1\!/_2)\right)
\label{19}
\end{eqnarray}
so that the average position of the $n$-th monomer is given by (only the component of $\langle \vec{R}_n(t)\rangle$ in the direction of the force is non-zero, so we drop vector notation)
\begin{eqnarray}
\langle R_n(t)\rangle= \frac{ft^\alpha}{\gamma N G_\alpha \alpha}Ê\left[1 + 2 \Gamma(\alpha +1) \sum_{p=1}^{N-1} \cos\left(\frac{\pi p}{2N}\right) \cos\left(\frac{\pi p}{N}(n+{}^1\!/_2)\right) E_{\alpha,\alpha+1} \left( - \left(\frac{t}{\tau_{p,\alpha}}\right)^\alpha \right)\right] \nonumber \\
\label{20}
\end{eqnarray}
from which we get the average velocity of the $n$-th monomer
\begin{eqnarray}
\langle V_n(t)\rangle= \frac{ft^{\alpha-1}}{\gamma N G_\alpha}Ê\left[1 + 2 \Gamma(\alpha) \sum_{p=1}^{N-1}\cos\left(\frac{\pi p}{2N}\right) \cos\left(\frac{\pi p}{N}(n+{}^1\!/_2)\right) E_{\alpha,\alpha} \left( - \left(\frac{t}{\tau_{p,\alpha}}\right)^\alpha \right)\right] 
\label{21}
\end{eqnarray}
From (\ref{20}) we can get information on the shape of the polymer as a function of time, while from (\ref{21}) we get insight into the velocity response of the polymer to turning on the force at $t=0$. This will allow us to determine how the tension propagates along the chain, a question recently studied for a polymer in a viscous environment \cite{Sakaue12,Rowghanian12}. We now discuss these issues in more detail.
\section{Shape of the polymer}
The applied force breaks the rotational symmetry of the polymer and it gets a length $L$ in the direction of the force. Moreover its density depends on the distance (measured along the direction of the force) from the first monomer. For a polymer in an ordinary viscous medium this leads to a trumpet shape, similar to that of a polymer in a constant flow \cite{Brochard93}. Since the Rouse chain has no finite extensibility, the stem-flower shape \cite{Brochard95} does not show up in this model. We do not expect these properties to change in a viscoelastic medium since they are equilibrium properties. What does change is the transient behaviour of these quantities. 

The (average) length $L(t)$ of the polymer is defined as
\begin{eqnarray}
L(t) = |\langle  \vec{R}_0(t) - \vec{R}_{N-1}(t)\rangle |
\label{22}
\end{eqnarray}
which using (\ref{20}) gives 
\begin{eqnarray}
L(t)   =\left| \frac{4  f t^\alpha}{\gamma N \Gamma(3-\alpha)} \sum_{p=1}^{N-1}  \cos^2 \left(\frac{\pi p}{2N}\right) \sin^2\left(\frac{\pi p}{2}\right) E_{\alpha,\alpha+1} \left( - \left(\frac{t}{\tau_{p,\alpha}}\right)^\alpha \right) \right|
\label{23}
\end{eqnarray}
Asymptotically in time, using (\ref{15})
\begin{eqnarray}
L^\star = L(t \to \infty)  = \frac{4 f N}{k \pi^2} \sum_{p=1}^{N} \sin^2\left(\frac{\pi p}{2}\right) /p^2 = \frac{fN}{2k}
\label{24}
\end{eqnarray}
The last equality holds again for $N\gg 1$. This is the same steady state length as for a Rouse chain pulled through a viscous medium \cite{Sakaue12}. This is no surprise, since the steady state is just the equilibrium state that only depends on temperature (enforced through the use of the fluctuation-dissipation theorem). 

However, the transient behaviour is different in the viscoelastic case. For times, where $\tau_{N,\alpha} \ll t$, the high $p$-modes have decayed so that the sum in (\ref{23}) can be approximated by the integral
\begin{eqnarray*}
\int_0^\infty \sin^2 \left(\frac{\pi p}{2}\right) E_{\alpha,\alpha+1} \left( - \frac{k \pi^2 p^2 t^\alpha}{\Gamma(3-\alpha) \gamma N^2}\right) dp
\end{eqnarray*}
After a switch of variables $y=pt^{\alpha/2}/N$ this leads to a scaling form for $L(t)$
\begin{eqnarray}
L(t) = t^{\alpha/2} F\left(\frac{t}{\tau_{R,\alpha}}\right)
\label{25}
\end{eqnarray}
where $F(x)$ is constant for $x$ small and goes as $x^{-\alpha/2}$ for $x \gg 1$. The time scale $\tau_{R,\alpha}$ is the natural extension of the Rouse time scale to the viscoelastic environment. For $N \gg 1$
\begin{eqnarray}
\tau_{R,\alpha} = \left[\frac{\Gamma(3-\alpha) \gamma N^2}{k \pi^2}\right]^{1/\alpha}
\label{26}
\end{eqnarray}
The dependence on $N$ as $N^{2/\alpha}$ implies that for the small $\alpha$-values reported for cells ($0.1 < \alpha < 0.4$), this time scale can become quite large. 

\begin{figure}
\includegraphics[width=14.0cm]{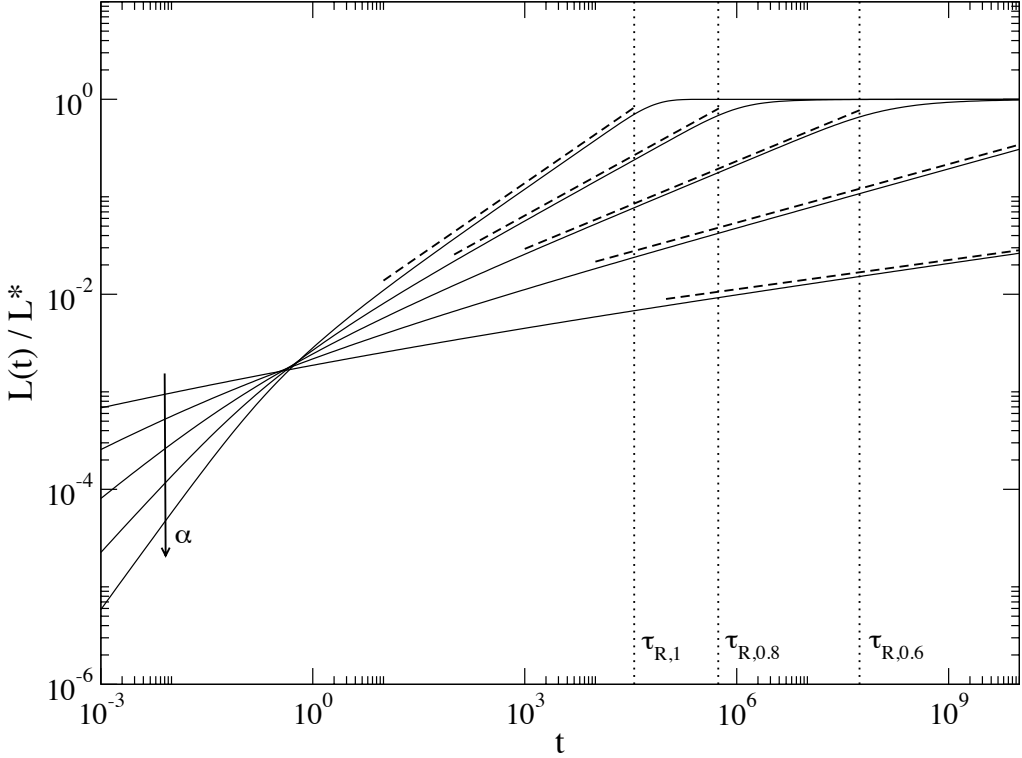}
\caption{\label{Fig1} Ratio of length to equilibrium length, $L(t)/L^\star$ , of a Rouse polymer subjected to a sudden force $f$ at $t=0$ as a function of time for $\alpha=0.2, 0.4, 0.6, 0.8$ and $1.0$ (increasing value of $\alpha$ is indicated by the arrow). All figures are for $f=2$ and $N=1024$. We have indicated the generalised Rouse time (\ref{26}) for various $\alpha$-values by vertical lines (for $\alpha=0.2$ and $0.4$ these times lay outside the figure). The straight lines have slope $\alpha/2$. }
\end{figure}

In Fig. \ref{Fig1} we plot $L(t)/L^\star$ as a function of time for various $\alpha$-values. We clearly see the predicted behaviour: after an initial regime, the length grows as $t^{\alpha/2}$ and then reaches its limiting value after $\tau_{R,\alpha}$. 

Next we determine the time evolution of the density profile. We define $X(n)$ as the average distance (in the direction of the force) between the $n$-th and the first monomer. One finds 
\begin{eqnarray*}
X(n,t) &=& \left| \langle \vec{R}_0(t)\rangle - \langle \vec{R}_{n}(t)\rangle \right| \nonumber \\
&=& \left| \frac{4ft^\alpha}{\gamma N \Gamma(3-\alpha)} \sum_{p=1}^N \cos\left(\frac{\pi p}{2N}\right)\sin\left(\frac{\pi p}{2N}(n+1)\right)\sin\left(\frac{\pi p}{2N}n\right) E_{\alpha,\alpha+1} \left( - \left(\frac{t}{\tau_{p,\alpha}}\right)^\alpha \right) \right|
\end{eqnarray*}
Inverting this relation gives $n(X,t)$ from which we obtain the density $\rho(X,t)=dn(X,t)/dX$. For arbitrary times this calculation has to be done numerically. However, in the steady state we can get analytical results. Indeed for $t \to \infty$ we obtain, using reasonings similar to those given for the equilibrium length $L^\star$ 
\begin{eqnarray}
X(n,t\to \infty) = \frac{f}{k} \left[ n - \frac{n^2}{2N} \right]
\label{28}
\end{eqnarray}
Inversion of this result and subsequent derivation gives
\begin{eqnarray}
\rho(X,t\to \infty) = \frac{k}{f \sqrt{1 - X/L^\star}}
\label{29}
\end{eqnarray}
This density profile, which diverges at $L^\star$ is the famous trumpet shape \cite{Brochard93} of a polymer in a steady flow. Notice that this result is again independent of the properties of the viscoelastic medium and therefore also holds for the original Rouse model (\ref{1}). As far as we know, this result was not derived before for that case. 

\begin{figure}
\includegraphics[width=14.0cm]{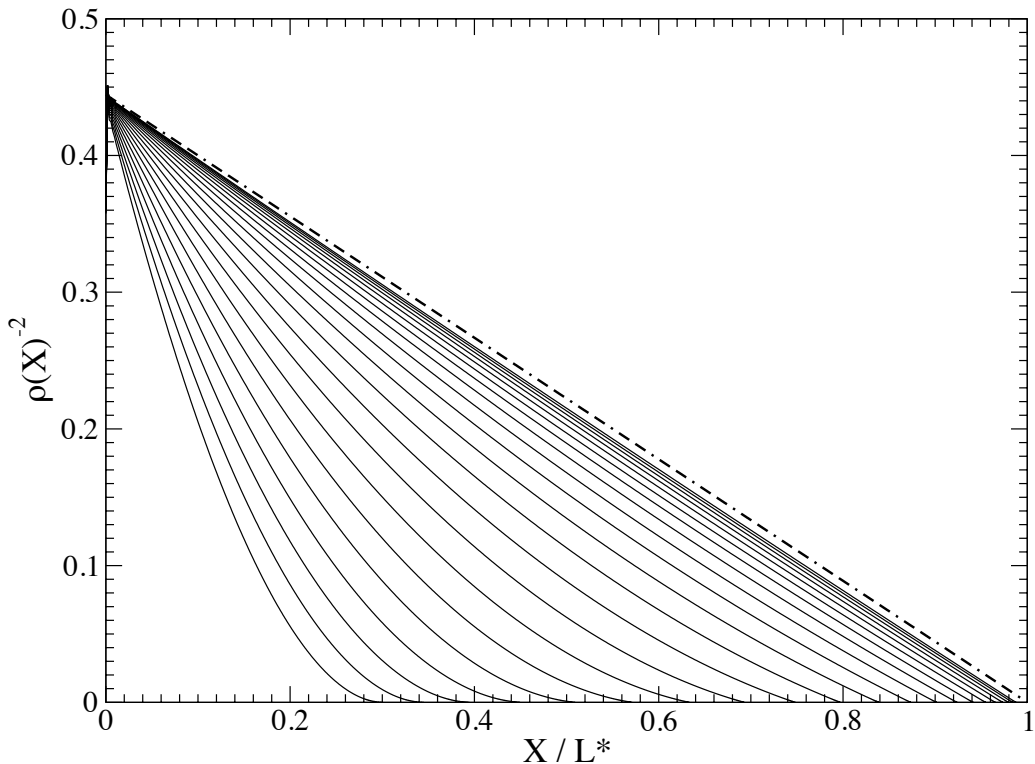}
\caption{\label{Fig2} Time evolution of the density $\rho(X)$ for a medium with $\alpha=0.4$. We plot $\rho(X)^{-2}$ versus $X$ for different times (increasing from left to right). The dash-dotted line gives the equilibrium shape (\ref{29}). The different curves give the density at times $t=2^i \tau_{R,\alpha}$ with $i=-6,-5,\ldots,13,14$ (left to right). }
\end{figure}

\begin{figure}
\includegraphics[width=14.0cm]{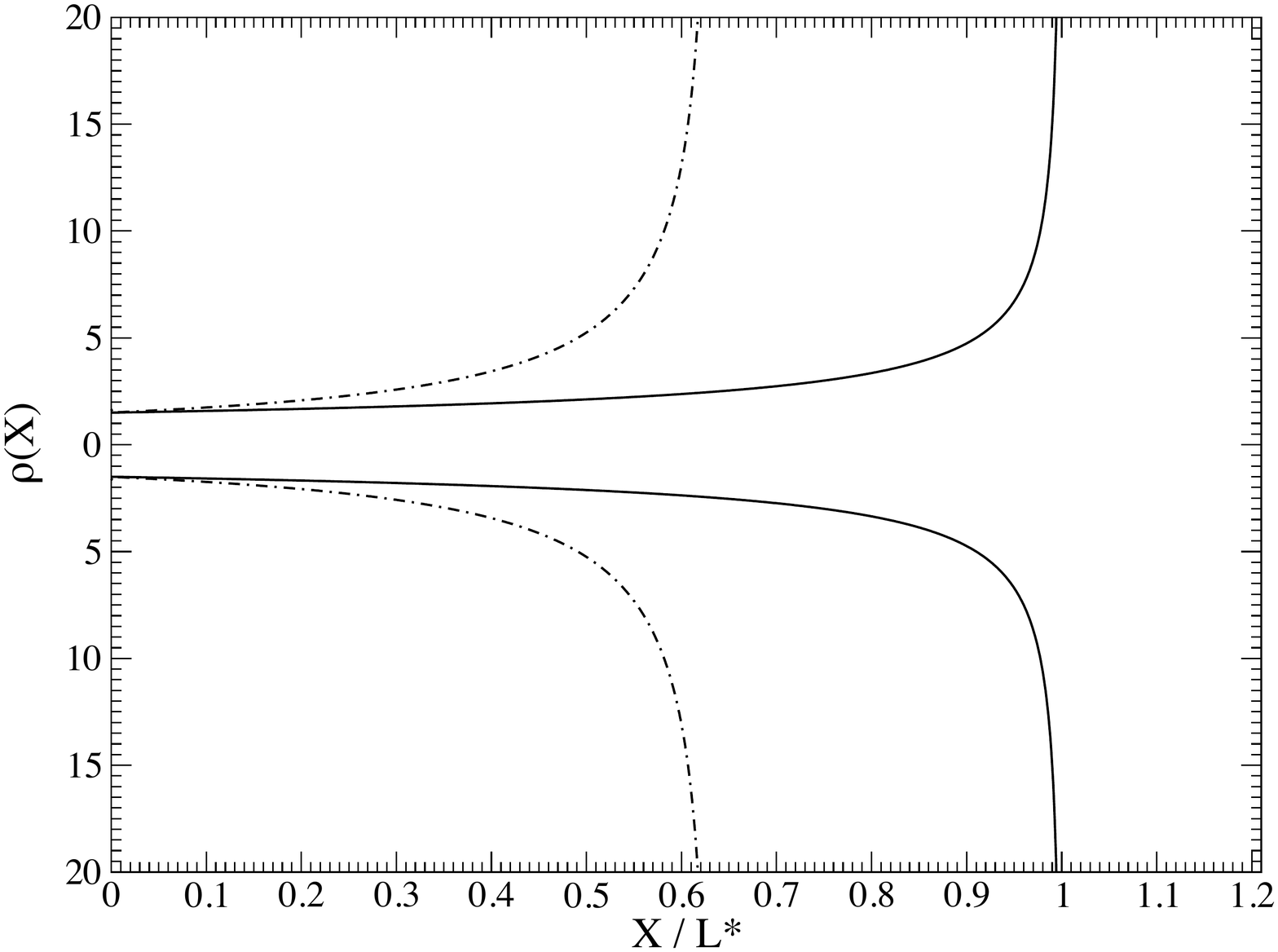}
\caption{\label{Fig3} Density of monomers versus distance from the first monomer for a medium with $\alpha=0.4$. The full line gives the equilibrium shape, the dashed-dotted one gives the density at the Rouse time $\tau_{R,0.4}$. The density is also shown reflected on the $X$-axis to show the trumpet shape. }
\end{figure}

Fig. \ref{Fig2} shows our (numerical) results for the density as a function of time for $\alpha=0.4$ and $N=1024$. Because of the form (\ref{29}) it is convenient to plot $\rho(X)^{-2}$ as a function of $X$ which in the steady state is a linear function that becomes zero at $L^\star$. For finite times, $\rho(X,t)$ diverges again at $L(t)$ though the precise function relation is complicated and we were not able to derive an analytical expression for it. Fig. \ref{Fig3} shows the trumpet shape of the polymer at $t=\tau_{R,\alpha}$ and in the steady state, again for $\alpha=0.4, f=2$ and $N=1024$. 

\section{Front propagation}
We next analyse the velocity of the monomers as given by (\ref{21}). Comparison with (\ref{13}) shows that the first term of (\ref{21}) is just the center of mass velocity. The second term, which we will denote by $J_n(t)=\langle V_n(t)\rangle - \langle V_{cm}(t)\rangle$, then denotes the difference in velocity between a given monomer and the center of mass. In the viscous case, $\alpha=1$, this term goes to zero exponentially fast after the Rouse time $\tau_{R,1}$. In the viscoelastic case, this transient term dies out much slower. Asymptotically in time, only the Rouse mode with $p=1$ survives, and then using (\ref{15b}) we find that $J_n(t)$ decays independently of $n$ as
\begin{eqnarray}
J_n(t \to \infty) \sim N^3 t^{-\alpha-1}
\label{29b}
\end{eqnarray}
If we look at  the first monomer ($n=0$), and approximate the sum in (\ref{21}) by an integral
\begin{eqnarray}
J_0(t) \approx \frac{2 f t^{\alpha-1}}{\gamma N \Gamma(3-\alpha)} \int_0^\infty \cos^2\left(\frac{\pi p}{2 N}\right) E_{\alpha,\alpha} \left( - \frac{k \pi^2 p^2 t^\alpha}{\Gamma(3-\alpha) \gamma N^2}\right) dp
\label{30}
\end{eqnarray}
we find that, after changing again to the variable $y$ defined in the previous section, that $J_0(t)$ is of the form
\begin{eqnarray}
J_0(t)= t^{\alpha/2-1} G\left(t^{\alpha/2}\right)
\label{31}
\end{eqnarray}
where $G(x)$ is a constant for small arguments, while $G(x) \sim x^{-3}$ for large arguments, in order to match (\ref{29b}). 

\begin{figure}
\includegraphics[width=14.0cm]{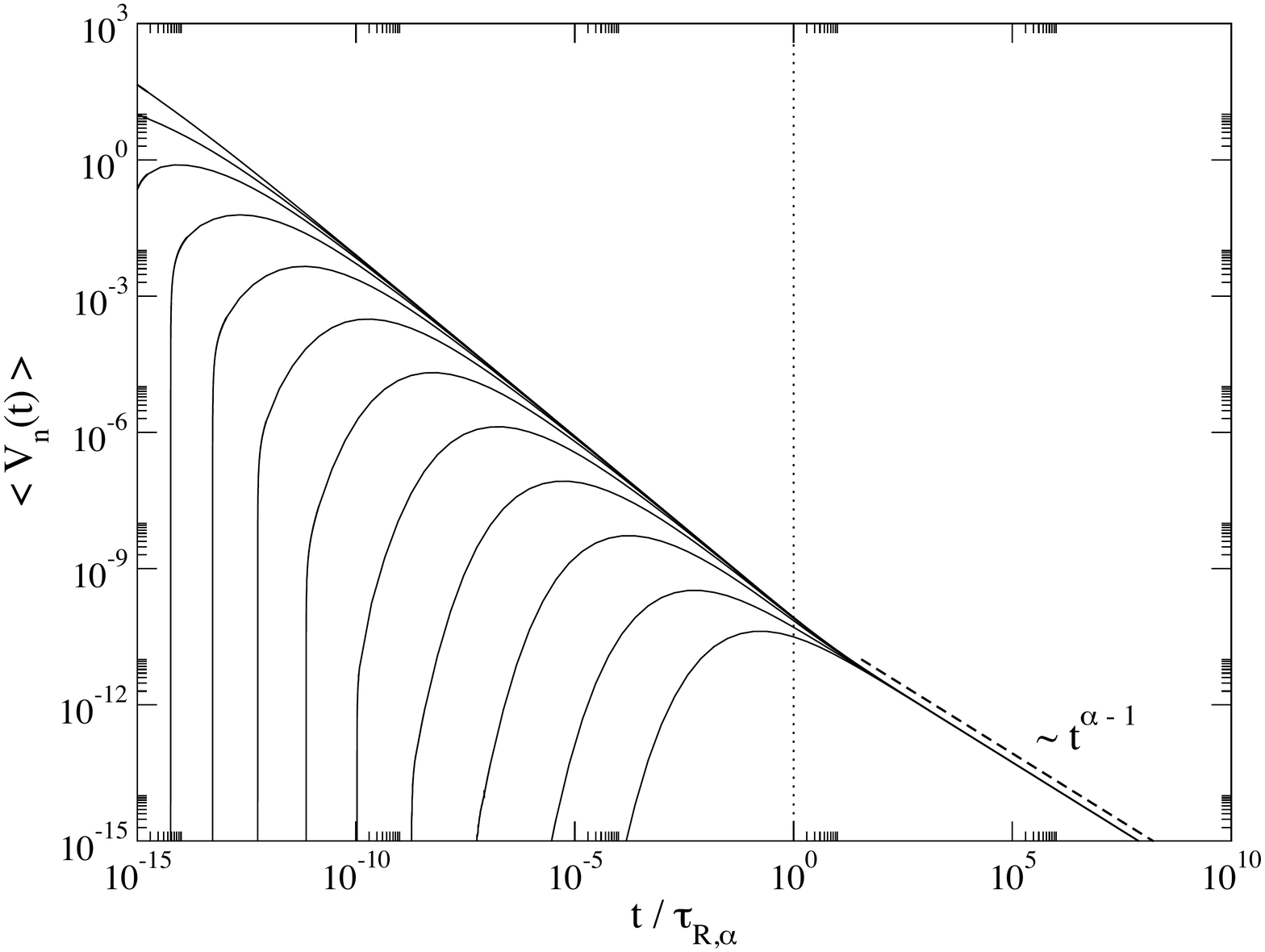}
\caption{\label{Fig4} Average velocity of the 'zeroth' monomer and $n$-th monomer versus time for $n=2^i, i=0,\ldots,9$ and $n=1023$ (left to right). The dashed line gives the late time behaviour for all monomers as derived in the text. The vertical line indicates the generalised Rouse time. Results are for $\alpha=0.4$.}
\end{figure}

\begin{figure}
\includegraphics[width=14.0cm]{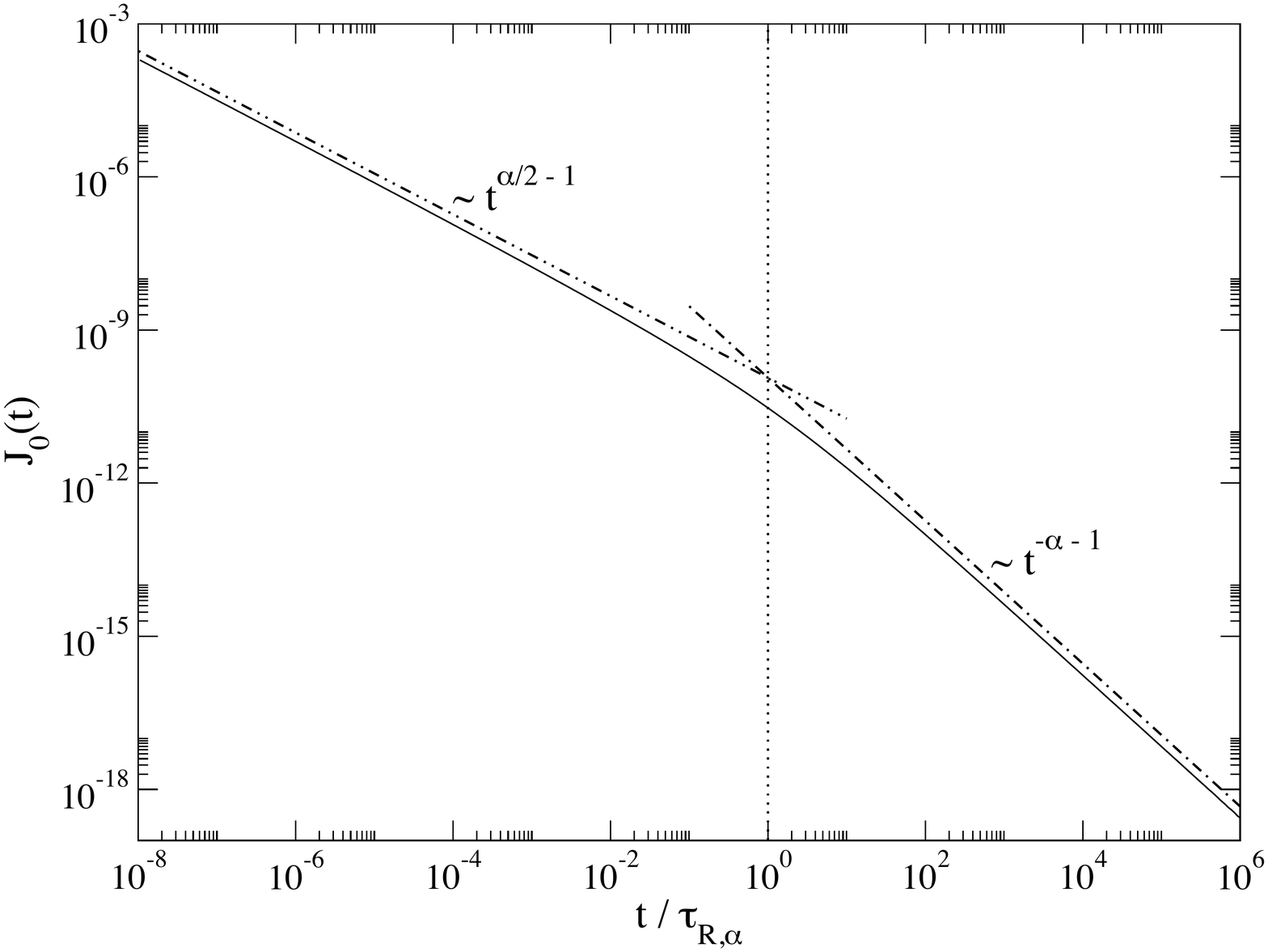}
\caption{\label{Fig5} Difference between the velocity of the first monomer and the velocity of the center of mass versus $t/\tau_{R,\alpha}$. Results are for $\alpha=0.4$. The dashed-dotted lines indicate the early and late time behaviour as discussed in the text.  }
\end{figure}

In Fig. \ref{Fig4} we plot the (average) velocity $\langle V_n(t)\rangle$ of a few monomers as a function of $t/\tau_{R,\alpha}$. The first monomer starts to move immediately (inertia is neglected in the Rouse model) and then as time goes on other monomers begin to move. After the Rouse time, all monomers move with the velocity of the center of mass, up to the power law corrections discussed above, which however cannot be seen on the scale of the figure. In Fig. \ref{Fig5} we show the exact results for $J_0(t)$. As predicted above, the initial decay $\sim t^{\alpha/2-1}$ crosses over to a faster decay $t^{-\alpha-1}$ after the Rouse time.  

\begin{figure}
\includegraphics[width=14.0cm]{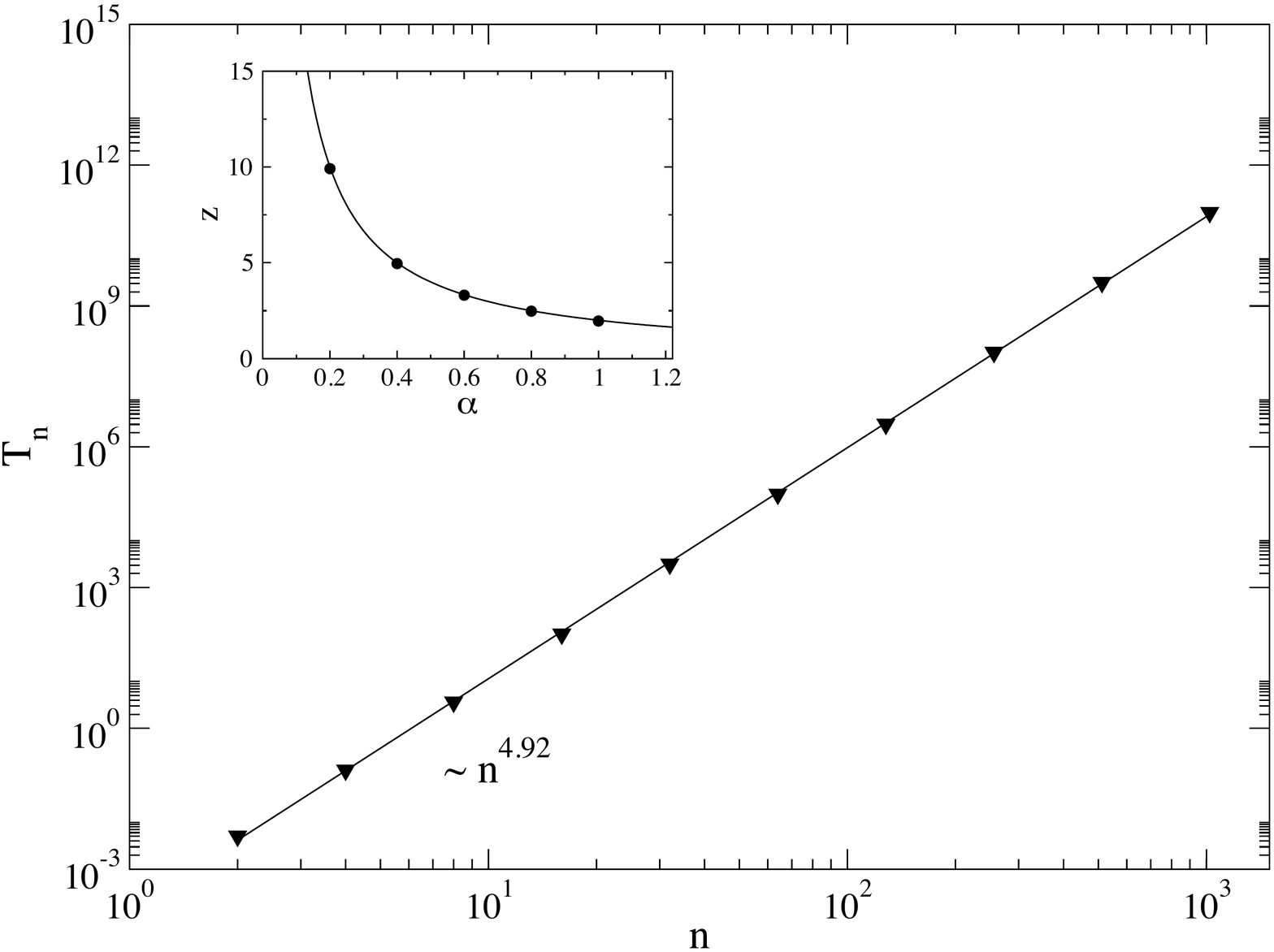}
\caption{\label{Fig6} Time $T_n$ at which the $n$-th monomer has a maximal velocity versus $n$ in a log-log plot, for a medium with $\alpha=0.4$. The straight line shows a best linear fit through the data points. The inset shows the estimates of the exponent $z$ versus $\alpha$. The full line is $z=2/\alpha$.}
\end{figure}

The data in Fig. \ref{Fig4} show that as we pull on the first monomer, a front moves through the polymer that gradually sets into motion all other monomers. In order to further characterise the motion of this front, we determined numerically the time $T_n$ at which the $n$-th monomer reaches a maximum velocity. One can say that at this time, the response to the pulling force is maximal. In Fig. \ref{Fig6} we plot these times as a function on $n$ in a log-log plot for $\alpha=0.4$. Clearly, the front propagates as a power law $T_n \sim n^z$. From a fit of the data, we find $z\approx4.92$. This estimate is consistent with the prediction $z=2/\alpha$ as can be expected from scalings like (\ref{25}). The inset of Fig. \ref{Fig6} shows that this expectation is indeed in agreement with the numerical results for the whole range of $\alpha$-values. 
\section{Fluctuations}
So far we calculated only the first moment of the position of each monomer, and quantities that can be derived from it. As shown in the appendix A, the monomer position is a Gaussian random variable and hence the position at a given moment will be fully determined if we also calculate its variance. This can be done though the calculations are lengthy. Here we focus on two properties that are interesting from a physical point of view: the fluctuations of the center of mass position, and the fluctuations in the polymer length.

It is convenient to introduce the variable $\vec{Y}_0(t)=\vec{X}_0(t)-\vec{X}_0(0)$, the distance travelled by the center of mass since the force was turned on. From (\ref{11}) we have 
\begin{eqnarray}
\langle Y^2_0(t) \rangle = \left(\frac{1}{\gamma G_\alpha}\right)^2 \int_0^t\int_0^t \langle \vec{F}_0(t-\tau)\cdot \vec{F}_0(t-\tau') \rangle \tau^{\alpha-1} \tau'^{\alpha-1} d\tau d\tau'
\label{32}
\end{eqnarray}
After using the expression (\ref{9}) for the force $\vec{F}_0(t)$ and the fluctuation-dissipation theorem (\ref{4}), we obtain
\begin{eqnarray}
\langle \vec{F}_0(t-\tau)\cdot \vec{F}_0(t-\tau') \rangle&=& \frac{1}{N^2} \sum_{n=0}^{N-1} \sum_{m=0}^{N-1} \left( 3 \gamma k_B T K(\tau-\tau') \delta_{nm} + f^2 \delta_{n0}\delta_{m0}\right) \nonumber \\
&=& \frac{1}{N^2} \left( \frac{3 \gamma N k_B T (2-\alpha)(1-\alpha)}{|\tau-\tau'|^\alpha} + f^2\right)
\label{33}
\end{eqnarray}
Inserting in (\ref{32}) gives
\begin{eqnarray}
\langle Y^2_0(t) \rangle = \left(\frac{1}{\gamma N G_\alpha}\right)^2\left[ \left(\frac{ft^\alpha}{\alpha}\right)^2 + 3 \gamma N k_B T (2-\alpha)(1-\alpha) \int_0^t\int_0^t \frac{\tau^{\alpha-1} \tau'^{\alpha-1}}{|\tau-\tau'|^\alpha} d\tau d\tau'\right]
\label{34}
\end{eqnarray}
The remaining integration can be done in terms of beta-functions and equals $2 t^\alpha \Gamma(\alpha) \Gamma(1-\alpha)/\alpha$. Combining the second moment of $\vec{Y}_0(t)$ with its average calculated in section III  then finally gives its variance (or fluctuation) as
\begin{eqnarray}
\sigma^2_{cm}(t) &\equiv& \langle Y^2_0(t) \rangle - \langle \vec{Y}_0(t)\rangle^2 \nonumber \\
&=& \frac{6k_BT}{\gamma N \alpha G_\alpha}t^\alpha
\label{35}
\end{eqnarray}
a result that, as could be expected, is independent of $f$ and which is in agreement with the result in \cite{Weber10} where the present model was studied in absence of a force. 

We finally look at the autocorrelation of the end-to-end vector $\vec{P}(t)=\vec{R}_0(t)-\vec{R}_{N-1}(t)$. The average size of this vector is the length of the polymer $L(t)$, see (\ref{22}). Using (\ref{19}), it is easy to show that for a large polymer
\begin{eqnarray}
\vec{P}(t) = 4 \sum_{p=1,odd}^N \vec{X}_p(t)
\label{36}
\end{eqnarray}
The autocorrelation of $\vec{P}(t)$ can therefore be rewritten in terms of the autocorrelations of the various modes $\vec{X}_p(t)$. These are calculated in appendix B. We find
\begin{eqnarray}
\langle \vec{P}(t)\cdot \vec{P}(t')\rangle &=& 16 \sum_{p=1,odd}^N \sum_{q=1,odd}^N \langle \vec{X}_p(t) \cdot \vec{X}_q(t') \rangle \nonumber \\
&=& \frac{16 f^2 t^\alpha t'^\alpha}{\gamma^2 N^2 \Gamma^2(3-\alpha)} \sum_{p=1,odd}^N \sum_{q=1,odd}^N  C_p C_q E_{\alpha,\alpha+1}\left(-\left(\frac{t}{\tau_{p,\alpha}}\right)^\alpha \right) E_{\alpha,\alpha+1}\left(-\left(\frac{t'}{\tau_{q,\alpha}}\right)^\alpha \right) \nonumber \\
&+& \frac{24 k_B T}{ \gamma N} \sum_{p=1,odd}^N \tau_p E_{\alpha,1}\left(-\left(\frac{|t-t'|}{\tau_{p,\alpha}}\right)^\alpha\right)
\label{37}
\end{eqnarray}
where $C_p=\cos(\pi p/2N)$.
From the asymptotic behaviour of the Mittag-Leffler functions one immediately finds that $\langle \vec{P}(t)\cdot \vec{P}(0)\rangle$ decays as $t^{-\alpha}$ in contrast with the much faster exponential decay in the viscous Rouse model. 
From (\ref{37}) we can also obtain the fluctuation in the length of the polymer
\begin{eqnarray}
\sigma^2_L \equiv \langle \vec{P}(t)\cdot \vec{P}(t)\rangle - L^2(t) 
\label{38}
\end{eqnarray}
The result is surprisingly simple
\begin{eqnarray}
\sigma^2_L = \frac{3 k_B T}{k} N
\label{39}
\end{eqnarray}
and can be understood from the equipartition theorem. Indeed, the random variables $\vec{R}_i - \vec{R}_{i-1}$ are independent in the Rouse model and appear in a quadratic way in the model's Hamiltonian. Since we also have that
\begin{eqnarray}
\vec{P}(t) = \sum_{i=1}^{N-1} \left(\vec{R}_{i-1}- \vec{R}_i\right),
\label{39bis}
\end{eqnarray}
(\ref{39}) follows from the equipartition theorem.

\section{Discussion and conclusions}
In this paper we have investigated transient properties of a Rouse polymer that moves through a viscoelastic environment under the influence of a force. The Rouse model is a too simple model to describe a real polymer. In realistic polymer models one has to include interactions between the monomers which at high temperatures swell the polymer. At low temperatures, below the theta temperature, the polymer collapses because of monomer-monomer attractions \cite{Vanderzande98}. The polymer then assumes a globular shape. Such a globular polymer can be used as a starting point to describe proteins. 

The steady state properties and the transient behaviour of polymers subjected to a force and moving through a viscous medium have been recently studied for such more realistic polymer models. Using scaling arguments \cite{Sakaue12} it has been argued that the front propagation is not influenced by self-avoidance or by hydrodynamic interactions, so that $T_n \sim n^2$ always holds for polymers at high temperatures. Since it is the tension propagation that determines the length of the polymer, this is also found to grow in the initial time regime as $t^{1/2}$, independently of self-avoidance and hydrodynamic interactions. We are currently investigating the predictions of \cite{Sakaue12} numerically \cite{Vandebroek14}.

It is difficult to extend the arguments of \cite{Sakaue12} to the viscoelastic case since they were based on the blob picture which cannot be easily extended to that situation. It could be that also in viscoelastic media,  the tension propagation is not influenced by self-avoidance since in the stretched region one expects self-avoidance to play a minor role anyhow.  Therefore, we expect the results $L(t) \sim t^{\alpha/2}$ to remain valid. This prediction should be verified numerically, but simulations for a polymer moving through a viscoelastic medium are difficult since one has to generate a large number ($N$) of long, independent histories of the correlated noise $\xi(t)$. 

It is nowadays experimentally possible to pull a polymer through a medium using optical tweezers. The shape of that polymer can be studied if it is made fluorescent \cite{Perkins95,Larson97}. In this way, it can in principle be possible to determine the length of the polymer as a function of time. From such measurements it would be possible to get an independent estimate of the exponent $\alpha$. This can be seen as a new active microrheological approach which complements existing approaches where $\alpha$ is determined from (passive) measurements of the subdiffusion of a particle immersed in the environment or from (active) measurements of the response to a periodic perturbation \cite{Robert10}. 
\\
\ \\

{\bf Acknowledgement} 
We thank Enrico Carlon for many useful discussions. 
\newpage
\appendix
\section{Time evolution of the Rouse modes}
In this appendix we give the solution to the equation of motion (\ref{7}) of the Rouse modes. This equation can be solved by Laplace transformation which gives
\begin{eqnarray}
\tilde{K}(s) \left[s \tilde{X_p}(s) - X_p(0)\right]= - \frac{\tilde{X_p}(s)}{\tau_p} + \frac{\tilde{F_p}(s)}{\gamma}
\label{A1}
\end{eqnarray}
where we denote the Laplace transform of a function $f(t)$ by $\tilde{f}(s)$. Equation (\ref{A1}) holds for any component of $\vec{X}_p$. 
The Laplace transform of the kernel equals $\tilde{K}(s)=\Gamma(3-\alpha) s^{\alpha-1}$. Inserting this in (\ref{A1}), gives after some elementary algebra
\begin{eqnarray}
\tilde{X_p}(s)= X_p(0) s^{-1}\left[1+ \frac{s^{-\alpha}}{\Gamma(3-\alpha)\tau_p}\right] ^{-1}+ \frac{\tilde{F}_p(s)}{\gamma \Gamma(3-\alpha)} s^{-\alpha} \left[1 + \frac{s^{-\alpha}}{\Gamma(3-\alpha)\tau_p}\right] ^{-1}
\label{A2}
\end{eqnarray}
Laplace transforms of the type appearing in (\ref{A2}) can be inverted using the integral \cite{Haubold11}
\begin{eqnarray}
\int_0^\infty e^{-st} t^{\beta-1} E_{\alpha,\beta}(at^\alpha)dt=s^{-\beta}\left(1-as^{-\alpha}\right)^{-1}
\label{A3}
\end{eqnarray}
This expression can be immediately applied to invert the first term of (\ref{A2}) while in the second term we recognise the Laplace transform of the convolution of the force $F_p(t)$ with a Mittag-Leffler function.
This leads to the result already given in (\ref{15c})
\begin{eqnarray}
\vec{X}_p(t) = \vec{X}_p(0) E_{\alpha,1}\left(- \left(\frac{t}{\tau_{p,\alpha}}\right)^\alpha\right) + \frac{1}{\gamma \Gamma(3-\alpha)} \int_0^t \vec{F}_p(t-\tau) \tau^{\alpha-1} E_{\alpha,\alpha} \left(- \left(\frac{\tau}{\tau_{p,\alpha}}\right)^\alpha\right) d\tau \nonumber \\
\label{A4}
\end{eqnarray}
Using (\ref{7}) it can be seen that $\vec{X}_p(t)$ can be expressed as a (complicated) linear combination of the Gaussian random variables $\xi_n(t)$. Hence it is itself Gaussian. 
\section{Correlation of the Rouse modes}
In order to calculate the correlation $\langle \vec{X}_p(t) \cdot  \vec{X}_q(t') \rangle$ we use a calculational approach first clearly outlined by Pottier \cite{Pottier03}. From (\ref{A2}) we can obtain an expression for $\tilde{X}_p(s) \tilde{X}_q(s')$, which is then averaged over the initial conditions and histories of the noise. For the initial conditions we have
\begin{eqnarray}
\langle \vec{X}_p(0) \cdot \vec{X}_q(0) \rangle = \frac{3k_B T}{k_p} \delta_{p,q}
\label{A5}
\end{eqnarray}
This result follows from the equipartition theorem applied to the initial, equilibrium, condition. Here $k_p=2 \gamma N/\tau_p \approx 2 k \pi^2 p^2/N$. In this way, one obtains
\begin{eqnarray}
\langle \tilde{X}_p(s) \tilde{X}_q(s')\rangle &=& \frac{3 \Gamma^2(3-\alpha) k_B T}{k_p} \frac{s^{\alpha-1}s'^{\alpha-1}}{(\Gamma(3-\alpha) s^\alpha + 1/\tau_p)(\Gamma(3-\alpha)s'^\alpha + 1/\tau_q)}\delta_{p,q} \nonumber \\ &+& \frac{\langle \tilde{F}_p(s) \tilde{F}_q(s')\rangle}{\gamma^2 (\Gamma(3-\alpha) s^\alpha + 1/\tau_p)(\Gamma(3-\alpha)s'^\alpha + 1/\tau_q)}
\label{A6}
\end{eqnarray}
The autocorrelation of the noise in Laplace variables
\begin{eqnarray}
\langle \tilde{F}_p(s) \tilde{F}_q(s') \rangle = \left\langle \int_0^\infty \int_0^\infty e^{-st}e^{-s't'} F_p(t) F_q(t') dt dt' \right\rangle
\label{A7}
\end{eqnarray}
can be obtained from the definitions (\ref{5}) and (\ref{9})  and the autocorrelation of the noise (\ref{4}).
In this way, we find
\begin{eqnarray}
\langle F_p(t) F_q(t') \rangle = \frac{f^2}{N^2} \cos\left(\frac{\pi p}{2 N}\right) \cos\left(\frac{\pi q}{2 N}\right) + \frac{3 \gamma k_B T}{2N} K|t-t') \delta_{p,q}
\label{A8}
\end{eqnarray}
which after inserting in (\ref{A7}) gives
\begin{eqnarray}
\langle \tilde{F}_p(s) \tilde{F}_q(s') \rangle= \frac{f^2}{N^2} \frac{1}{ss'} \cos\left(\frac{\pi p}{2 N}\right) \cos\left(\frac{\pi q}{2 N}\right) + \frac{3 \gamma\Gamma(3-\alpha) k_B T}{2N} \ \left(\frac{s^{\alpha-1} + s'^{\alpha-1}}{s+s'}\right) \delta_{p,q} \nonumber \\
\label{A9}
\end{eqnarray}
Here we have used a general result on the double Laplace transform of functions that depend only on the absolute value of the time difference $|t-t'|$ (see eqn (2.16) in \cite{Pottier03}). Inserting this in (\ref{A6}) gives
\begin{eqnarray}
\langle \tilde{X}_p(s) \tilde{X}_q(s')\rangle &=& \frac{3 \Gamma^2(3-\alpha) k_B T}{k_p} \frac{s^{\alpha-1}s'^{\alpha-1}}{(\Gamma(3-\alpha) s^\alpha + 1/\tau_p)(\Gamma(3-\alpha)s'^\alpha + 1/\tau_q)}\delta_{p,q} \nonumber \\ &+& \frac{f^2}{\gamma^2 N^2} \cos\left(\frac{\pi p}{2 N}\right) \cos\left(\frac{\pi q}{2 N}\right) \frac{s^{-1} s'^{-1}}{(\Gamma(3-\alpha) s^\alpha + 1/\tau_p)(\Gamma(3-\alpha)s'^\alpha + 1/\tau_q)} \nonumber \\
&+& \frac{3 \Gamma(3-\alpha) k_B T}{2\gamma N} \ \frac{s^{\alpha-1}+s'^{\alpha-1}}{(s+s')(\Gamma(3-\alpha) s^\alpha + 1/\tau_p)(\Gamma(3-\alpha)s'^\alpha + 1/\tau_q)}\delta_{p,q} \nonumber \\ 
\label{A10}
\end{eqnarray}
Finally, we have to take the inverse Laplace transform of this equation. The second term is a product of a function in $s$ and one in $s'$ and the inverse Laplace transform can be expressed in terms of Mittag-Leffler functions using (\ref{A3}). The first and the third term can be added and give 
\begin{eqnarray}
\frac{3k_B T \tau_p}{2 \gamma N (s+s')} \delta_{p,q} \left( \frac{1}{s^{1-\alpha} (s^\alpha + 1/(\Gamma(3-\alpha)\tau_p))} + \frac{1}{s'^{1-\alpha} (s'^\alpha + 1/(\Gamma(3-\alpha)\tau_p))}\right)
\label{A11}
\end{eqnarray} 
which is of the form of the Laplace transform of a function of $|t-t'|$ \cite{Pottier03}. Putting everything together gives
\begin{eqnarray}
\langle X_p(t) X_q(t')\rangle &=& \frac{3 k_B T \tau_p}{2\gamma N} \delta_{p,q} E_{\alpha,1} \left(- \frac{|t'-t|^\alpha}{\Gamma(3-\alpha) \tau_p}\right) \nonumber \\
&+& \frac{f^2}{\gamma^2 N^2 \Gamma^2(3-\alpha)} C_p C_q t^\alpha t'^\alpha E_{\alpha,\alpha+1} \left(- \frac{t^\alpha}{\Gamma(3-\alpha) \tau_p}\right) E_{\alpha,\alpha+1} \left(- \frac{t'^\alpha}{\Gamma(3-\alpha) \tau_q}\right) \nonumber \\ 
\label{A12}
\end{eqnarray}
where $C_p=\cos\left(\frac{\pi p}{2 N}\right)$.

One easily verifies that in absence of a force the equipartition theorem holds at every time $\langle X_p^2(t)\rangle = 3 k_B T/k_p$.


\newpage

\begin{thebibliography}{100}
\bibitem{Doi86}
M. Doi and S.F. Edwards, {\it The theory of polymer dynamics}, Oxford University Press (1986).
\bibitem{Brochard93}
F. Brochard-Wyart, Europhys. Lett. {\bf 23}, 105 (1993).
\bibitem{Brochard95}
F. Brochard-Wyart, Europhys. Lett. {\bf 30}, 387 (1995).
\bibitem{Perkins95}
T.T. Perkins, D.E. Smith, R.G. Larson and S. Chu, Science {\bf 268}, 83 (1995).
\bibitem{Larson97}
R.G. Larson, T.T. Perkins, D.E. Smith and S. Chu, Phys. Rev. E {\bf 55}, 1794 (1997). 
\bibitem{Rowghanian12}
P. Rowghanian and A.Y. Grosberg, Phys. Rev. E {\bf 86}, 011803 (2012). 
\bibitem{Sakaue12}
T. Sakaue, T. Saito and H. Wada, Phys. Rev. E {\bf 86}, 011804 (2012).
\bibitem{Sakaue10}
T. Sakaue, Phys. Rev; E {\bf 81}, 041808 (2010).
\bibitem{Rowghanian11}
P. Rowghanian and A.Y. Grosberg, J. Phys. Chem. B {\bf 115}, 14127 (2011).
\bibitem{Panja08}
D. Panja and G.T. Barkema, Biophys. J. {\bf 94}, 1630 (2008). 
\bibitem{Boal12}
D. Boal, {\it Mechanics of the cell}, Cambridge University Press (2012).
\bibitem{Brangwynne08a}
C.P. Brangwynne, G.H. Koenderink, F.C. MacKintosh and D.A. Weitz, Phys. Rev. Lett. {\bf 100}, 118104 (2008). 
\bibitem{Howard01}
J. Howard, {\it Mechanics of motor proteins and the cytoskeleton}, Sinauer (2001). 
\bibitem{Lau03}
A.W.C. Lau, B.D. Hoffman, A. Davies, J.C. Crocker, and T.C. Lubensky, Phys. Rev. Lett. {\bf 91} 198101 (2003). 
\bibitem{Robert10}
D. Robert, T.-H. Nguyen, F. Gallet and C. Wilhelm, PLoS ONE {\bf 5}, e10046 (2010).
\bibitem{Hofling13}
F. H\"ofling and T. Franosch, Rep. Prog. Phys. {\bf 76} 046602 (2013). 
\bibitem{Qian03}
H. Qian, in {\it Process with Long-Range Correlations: Theory and Applications}, edited by G. Rangarajan and M.Z. Ding, Lecture Notes in Physics, vol. 621, Springer (2003). 
\bibitem{Fabry01}
B. Fabry {\it et al.}, Phys. Rev. Lett. {\bf 87} 148102 (2001).
\bibitem{Weber12}
S.C. Weber, A.J. Spakowitz and J.A. Theoriot, Proc. Natl. Acad. Sci. U.S.A. {\bf 109}, 7338 (2012). 
\bibitem{Weber10}
S.C. Weber, J.A. Theriot and A.J. Spakowitz, Phys. Rev. E {\bf 82}, 011913 (2010). 
\bibitem{Caspi00}
A. Caspi, R. Granek, and M. Elbaum, Phys. Rev. Lett. {\bf 85}, 5655 (2000). 
\bibitem{Haubold11}
H.J. Haubold, A.M. Mathai and R.K. Saxena, J. Appl. Math., 298628 (2011)
\bibitem{Vanderzande98}
C. Vanderzande, {\it Lattice models of polymers}, Cambridge University Press (1998). 
\bibitem{Vandebroek14}
H. Vandebroek, E. Carlon and C. Vanderzande, unpublished.
\bibitem{Pottier03}
N. Pottier, Physica A {\bf 317} 371 (2003). 


\end{thebibliography}
\end{document}